# H2-Golden-Retriever: Methodology and Tool for an Evidence-Based Hydrogen Research Grantsmanship

Paul Seurin[1,*], Olusola Olabanjo[2,*], Joseph Wiggins[3,*], Lorien Pratt[4], Loveneesh Rana[5], Rozhin Yasaei[6], Gregory Renard[7]

[1]Massachusetts Institute of Technology, 77 Massachusetts Avenue, Cambridge, MA 02139

[2]Morgan State University, 1700 E Cold Spring Ln, Baltimore, MD 21251

[3]Katabasis, Inc., 900 Riverside Rd., Grifton, NC 28530

[4]Quantellia LLC, 100 S. Murphy St Suite 20 Sunnyvale, CA 94086

[5]University of Luxembourg, 6 Rue Richard Coudenhove Kalergi L 1359, Walferdange

[6]University of California Irvine, 260 Aldrich Hall Irvine, CA 92697-1075

[7]AAICO, 743 Upload Road, Redwood City, California, 94062

⋆These authors who have contributed equally to this manuscript

## Abstract

**Background of Study:** Hydrogen is poised to play a major role in decarbonizing the economy. The need to discover, develop, and understand low-cost, high-performance, durable materials that can help maximize the cost of electrolysis as well as the need for an intelligent tool to make evidence-based Hydrogen research funding decisions relatively easier warranted this study.

**Aim:** In this work, we developed H2 Golden Retriever (H2GR) system for Hydrogen knowledge discovery and representation using Natural Language Processing (NLP), Knowledge Graph and Decision Intelligence. This system represents a novel methodology encapsulating state-of-the-art technique for evidence-based research grantmanship.

**Methods:** Relevant Hydrogen papers were scraped and indexed from the web and preprocessing was done using noise and stop-words removal, language and spell check, stemming and lemmatization. The NLP tasks included Named Entity Recognition using Stanford and Spacy NER, topic modeling using Latent Dirichlet Allocation and TF-IDF. The Knowledge Graph module was used for the generation of meaningful entities and their relationships, trends and patterns in relevant H2 papers, thanks to an ontology of the hydrogen production domain. The Decision Intelligence component provides stakeholders with a simulation environment for cost and quantity dependencies. PageRank algorithm was used to rank papers of interest.

**Results**: Random searches were made on the proposed H2GR and the results included a list of papers ranked by relevancy score, entities, graphs of relationships between the entities, ontology of H2 production and Causal Decision Diagrams showing component interactivity. Qualitative assessment was done by the experts and H2GR is deemed to function to a satisfactory level.

**Conclusion**: This work demonstrated that NLP and human-in-the-loop AI have a great potential for faster knowledge discovery. Also, using an ontology based on hydrogen production enables to uncover papers that would not be emphasized with traditional citation-based metrics. This tool is promising to reduce the workload of experts in searching and traversing the enormous H2 papers space that is released every day.

## I.    Introduction

Hydrogen has long been touted as a clean, versatile energy carrier of the future and could facilitate clean-energy pathways across multiple applications and sectors such as transportation, power generation, grid-scale storage of electricity (i.e power-to-gas) [1], food industry to harden animal or vegetable fats, or even ammonia and methanol production [2]. Nevertheless, hydrogen must be produced without carbon emissions to satisfy this goal. Water electrolysis powered by renewables or nuclear satisfies this constraint, but its adoption has been inhibited by technological and cost requirements. Typically Steam Methane Reforming (SMR) costs about 2-3 times less than water electrolysis with a price of hydrogen from electrolysis estimated around 5-6$/kg at an electricity cost of 0.05-0.07 $/kWh [3]. Hence, only 3.9% is produced with this method (water electrolysis) against 48% from natural gas reforming (Blue Hydrogen), 30% as a by-product of oil/naphtha reforming, 18% from coal gasification (Grey Hydrogen) and 0.1% from other sources [4].

The 2015 Paris agreement accentuated the urge for global cooperation between the key economic stakeholders, including the United States of America (USA), to mitigate the ever-accelerating climate change[5]. The Biden-Harris administration renewed its commitment toward this goal and pledged to achieve net-zero carbon emission by 2050. Hydrogen production could be a key factor in helping decarbonize the economy while boosting it and generating hundreds of thousands of high-paying jobs. In 2020, more than 20 countries presented their hydrogen strategies. In 2021, President Biden signed the bipartisan infrastructure law, which included 9.5 billion dollars for clean hydrogen, 1 billion dollars of which is dedicated to electrolysis research, development, and demonstration. The USA's Department of Energy (DOE) Hydrogen Energy Earthshot established a landmark goal to advance hydrogen production technologies to produce 1 kg H2 for 1$/kg in the next decade (coined the "1 1 1" goal) and 2 $/kg by 2026.

The electrolyzer system for hydrogen production comprises the balance-of-plant (power supplies, hydrogen processing, cooling, etc.) and the stack (where water is split into hydrogen and oxygen)[6]. Significant cost reductions can be realized through economies of scale for balance-of-plant and stack components, particularly for power electronics, but these cost reductions are not sufficient[7]. R&D is vital to further reduce the stack cost, representing about one-third of the production cost (where the most significant contributors to the cost are manufacturing process yield, power density, Platinum Group Metal (PGM) loading, and the membrane) [7]. Many researchers across the globe are working to discover, develop, and understand low-cost, high-performance, durable materials that can help bring down the cost of electrolysis. Researchers have previously utilized published materials data to model behavior and identify promising materials for hydrogen storage[8]. However, unlike hydrogen storage materials, electrolyzers include multiple components, which do not produce numerical data in standardized ways that can be gathered in standardized databases. On top of the plethora of papers published in the field of hydrogen production (which exceeded 100,000 since 2021 based on a Google Scholar keyword search), researchers could benefit from other broader domains such as the advancement in fuel cell design[9] ionically conducting membranes[10], ion exchange membrane [8], or membrane for water desalination[11]. This massive amount of research is challenging to track even by human experts, who are, at the time of this research, still navigating the research space manually. They could therefore benefit from intelligent tools that help them identify materials and methods most valuable to them.

On top of that, while decision-makers are not conducting research directly, they are interested in funding advanced science approaches that may help reduce the hydrogen cost vis-à-vis efficient production, availability and distribution. The current - manual - knowledge discovery method involves the use of publication repository such as Google scholar to receive weekly alerts on relevant papers based on keyword searches and the use of expertise to select the most relevant based on the title, the references/citations, and the abstract. Technical articles of interest are then read to find relevant information and further insights to orient their keyword searches in a more promising direction. This makes keeping up with the torrential pace of publications very difficult, cumbersome, time-consuming and labor-intensive.

Since Hydrogen production cost depends on multiple assumptions including materials, electrolyzer design, manufacturing practices, and soft costs, it is difficult to trace back the contribution of the price of a component to the overall system cost. This makes cost of a component to vary from one technical report to another; for instance, a study[12] reported that bipolar plates is the third contributor behind the Catalyst Coated Membrane (CCM) and Porous Transport Layer (PTL) to the stack cost while another article[13] reported it as the highest contributor. It is therefore hard to crystallize on which sub-component deserves the highest focus. Lastly, experts are on a race against time. Understanding how one component affects the overall cost is one thing but how much money must be invested in versus the time it will take to reduce the overall system cost must drive the decisions of the experts. In this work, we addressed the missing link between the slew of information available in literature and the decision to fund research and, by implication, propose an AI-augmented decision tool for knowledge acquisition, knowledge extraction and an evidence-based research funding decision support tool.

## II. $H_2$ Production Methods

Today about 48% of hydrogen is used for ammonia manufacturing, 37% for petroleum refineries, 8% for methanol production and the reminder for smaller volume application[14]. Nevertheless, hydrogen must be produced without carbon emissions (product of which is referred to as Green $H_2$) to satisfy this goal. Water electrolysis powered by renewable or nuclear satisfies this constraint, but its adoption has been inhibited by technological and cost requirements[13]. In spite of producing a less pure hydrogen and producing harmful and greenhouse gases[15], Steam Methane Reforming (SMR) costs (usually evaluated around 2$/kg) about 2-3 times less than from water electrolysis with a price of hydrogen from electrolysis estimated around 5-6$/kg at an electricity cost of 0.05-0.07 $/kWh[3]. Hence, out of the 368 trillion cubic meters of hydrogen produced every year [50], only 3.9% is produced with this method (water electrolysis) against 48%from natural gas reforming (Blue Hydrogen), 30% as a by-product of oil/naphtha reforming, 18%from coal gasification (Grey Hydrogen) and 0.1% from other sources[14].

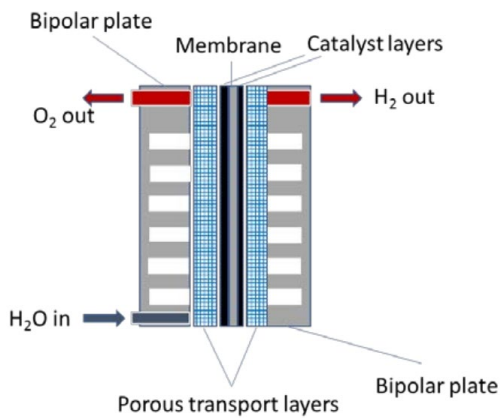

*Figure 1: Schematic of a PEM electrolyzer (picture taken from [12]).*

There are four types of water electrolyzer: The Alkaline, The Proton Exchange Membrane (PEM, which schematic is given in figure 1.), the Anion Exchange Membrane (AEM), and the Solid Oxide electrolyzer (SOEC)[13]. Only the first two reached technological maturity and are today produced at commercial scale, while the last two are produced at lab scale and promise further headway in reducing the cost of green hydrogen. In the alkaline electrolyzer, the electrolyte is liquid in the form of KOH. This is the main technology utilized today as it is less expensive than its PEM counterpart which has a 50-60%higher cost[13] (or evaluated at 3-5 C/kg[16]). This is due, in particular, to expensive metals used in the acidic environment at the anode of the PEM[11]. Nevertheless, Alkaline electrolyzers are limited by low current density, larger footprint, and are less adapted to load variation[3], preventing them to be coupled with renewable for power-to-gas application. On the other hand, the PEM electrolyzer is characterized by a high current density (over 2 A/cm2), high Faradaic efficiency, fast response time[17], utilize pure water instead of the corrosive KOH-based electrolyte, and can operate in differential pressure thanks to the mechanically and chemically robust membrane (e.g., Perfluorosulfonic Acid (PFSA)) (it can operate up to 70 bar with the anode at atmospheric pressure simplifying the oxygen side of the balance of plants and reducing the load on hydrogen compression requirement, enhancing safety and cost in the process).

The PEM electrolyzer system for hydrogen production comprises the balance-of-plants (power supplies (DC rectifier), hydrogen compression, hydrogen cooling, water deionization and purification) and the stack (where water is split into hydrogen and oxygen). Studies are in accordance to say that the balance of plant contributes to about 55% of the system cost and the stack the remaining 45%[13]. A significant cost reduction potential resides in the economy of scale and balance-of-plant (in particular the power supply) because the electrolyzer is greatly affected by the price of electricity[3], but this is insufficient[12]. R&D is vital to reduce the stack cost, and the provenance of energy sources depends on where the hydrogen is produced. Therefore, the following discussion focuses on the stack. The stack is made of four main components namely the bipolar plates (BPPs), the Catalyst Coated Membrane (CCM), the Porous Transport Layers (PTLs), and other parts including the end plates and seal frames. However, for cost evaluation, the balance-of-stacks and the stack assembly line are usually included. The BPPs, structured with a flow channel, ensure mechanical support and even distribution of the flow going in and out of the cell. The CCM, sandwiched between the PTLs, is made of the catalysts coated onto the polymer electrolyte (i.e the membrane). The PTLs ensure thermal and electrical conductivity while facilitating the transport of the reactants and the removal of products. Water is usually introduced at the anode, transported through the anodic PTL, and then reaches the anodic catalyst layer where the Oxygen Evolution Reaction (OER, equation 2) takes place. The oxygen is then removed from the cell via the anodic PTL. The Proton created during OER flows through the membrane and is reduced to hydrogen via the Hydrogen Evolution Reaction (HER, equation 1, 2 and 3) at the cathodic catalyst layer. Water is also transported through the membrane by electro-osmotic drag. The hydrogen is then transported through the cathodic PTL and removed from the cell. The end plates and seal frames provide mechanical support, prevent leakage, and collect the reactants [22].

$$H_2O \rightarrow 2H^+ + \frac{1}{2}O_2 + 2e^- \quad (1)$$

$$2H^+ + 2e^- \rightarrow H_2 \quad (2)$$

The global reaction is

$$H_2O \rightarrow H_2 + \frac{1}{2}O_2 \quad (3)$$

### III.    H₂ Ontology

Domain information and knowledge representation are important in the categorization and annotation of products, standardize domain semantics, share and re-use information in a specific domain[18] to facilitate scientific collaboration, and analyzing the domain knowledge. In this perspective, the notion of ontology, which originates from the semantic web stack [19], is a technology that can help in this endeavor. For instance, the classical application of ontology lies in the World-Wide Web where searching agents can seek information thanks to large, organized, taxonomies of websites, products, and features.

An ontology can be defined as a semantic network that maps concepts (or classes) and their relationships (or predicates) in a domain [20] abstracting common information into domain knowledge. The relations are classically represented in the form of a triplet subject-predicate-object, which can then be processed and understood by a computer. The WWW Consortium (W3C) has developed a unified language, the Resource Description Framework (RDF) [21], to encode the ontology into a machine-readable format. Typically, within the context of water electrolysis, general classes can be defined (e.g., PEM electrolyzer, membrane, and catalyst), which are associated to each other with predicates (e.g., *has a* would result in PEM electrolyzer *has a* catalyst, PEM electrolyzer *has a* membrane). Slots (or properties/attributes) are also assigned to classes, which characteristics (or facets) such as the range, value type, or cardinality can be defined to augment the expressiveness of the ontology [18]. Instances then populate the ontology (e.g. SIEMENS electrolyzer has a NAFION membrane, SIEMENS electrolyzer has a iridium-based catalyst), which gives rise to Knowledge Graphs (KG).

It is common, before releasing an ontology, to cooperate both with domain and ontology experts [22]. We hope to enhance the current ontology with more experts in hydrogen's production and ontology domain later on. However, due to the complexity of defining objective logical rules and complex physical properties, this preliminary work does not specify particular facets or axioms but only concepts and relationships.

## IV. Contributions of Study

This research proposed integrated, state-of-the-art, Artificial Intelligence (AI) tools which are encapsulated in a friendly User-Interface (UI) to facilitate the knowledge discovery, extraction and processing tasks to evidence-based research grantmanship. These tools include:

i. Natural Language Processing (NLP): to review the large corpus of related hydrogen research and extract important entities and topics rather than only meta-data (e.g., keywords, number of citations...). It will leverage semantic web technologies (e.g., ontology of hydrogen production) to make inferences and draw connections between papers.
ii. Ontology: An ontology is a representation of concepts and their relationship within a domain. The ontology of the hydrogen water electrolysis will be built and input to the NLP to generate a knowledge graph.
iii. Knowledge Graph (KG): The entities extracted will be input in a Knowledge Graphs (KG). This provides insights into the state of hydrogen science, uncovers relationships between research alleys, and discovers promising research trends. In turn, finding the appropriate papers with less time and effort.
iv. Paper Ranking: From the KG, a domain-augmented ranking strategy will be applied to suggest promising research direction within the domain of interest (hydrogen production in particular PEM electrolysis). This is to our knowledge the first attempt to apply an automatic paper ranking strategy incorporating domain knowledge and not only meta-data (e.g. citation, co-authors, etc…).
v. A Graph-based based smart recommender system: that utilizes the KG and a hydrogen domain-augmented score to suggest promising research direction.
vi. A Causal Decision Diagram (CDD). A CDD is a classical tool leveraged in Decision Intelligence[23] to reflect the decision mental models of the decision maker (e.g., hydrogen funding choices), which connects the dots between available technologies (e.g., cost) back to human's natural ways of thinking (e.g., what investment to make to advance towards the "1 1 1" goal).

### V. Materials and Methods:

In this work, we propose H2 Golden Retriever (H2GR): a tool developed to assist Hydrogen experts in discovering, uncovering and predicting research trends in green H2 production and serve as a machine-learning-assisted Decision Support System for experts in making funding decisions. It is an integrated tool for an easier and effective way for searching, mining, monitoring, discovery and analysis of relevant H2 production papers, especially with respect to cost and funding. H2GR provides a platform for performing a semantic search, applying interactive filters and data visualizations for searching, discovery and analysis of H2 papers for analysis, structuring, filtering and visualization of search results from a real-time corpus of relevant H2 papers. It is a powerful tool which automatically indexes papers on H2 production with special focus on funding directions and production costs. In this section, the materials and methods of implementing the proposed system are discussed.

#### a. Data Description

The papers that we utilized to conduct the search were found by keyword searches on Science Direct. In particular, the keyword utilized were along the line of "membrane cost electrolyzer", "water electrolysis", "cost catalyst", or "ion exchange membrane", which amounted to a total of 1573 papers. The journals used for this study were Applied Energy, International Journal of Hydrogen Energy (IJHE), Journal of Membrane Science, Catalysis Today, Electrochemica Acta, Journal of The Electrochemical Society, Applied Catalysis B: Environmental, RRL Solar, Energy Conversion and Management, Journal of Materials Science Technology, Progress in Natural Science: Materials International, and Joule among others. Some of the papers scraped from these journals are given in Figure 2

| Paper Title |
| --- |
| 1D two-phase, non-isothermal modeling of a proton exchange membrane water electrolyzer: An optimization perspective |
| A comprehensive modeling method for proton exchange membrane electrolyzer development |
| A data-driven digital-twin model and control of high temperature proton exchange membrane electrolyzer cells |
| A green hydrogen energy storage concept based on parabolic trough collector and proton exchange membrane electrolyzer/fuel cell: Thermodynamic and exergoeconomic analyses with multi-objective optimization |
| Analysis of the effect of characteristic parameters and operating conditions on exergy efficiency of alkaline water electrolyzer |
| An integrated system for ammonia production from renewable hydrogen: A case study |
| A novel approach to ammonia synthesis from hydrogen sulfide |
| A solid oxide membrane electrolyzer for production of hydrogen and syn-gas from steam and hydrocarbon waste in a single step |
| Assessment of power-to-power renewable energy storage based on the smart integration of hydrogen and micro gas turbine technologies |
| Assessment of thermal performance improvement of GT-MHR by waste heat utilization in power generation and hydrogen production |
| A stacked interleaved DC-DC buck converter for proton exchange membrane electrolyzer applications: Design and experimental validation |
| A study on sustainability analysis of solid oxide fuel cell and proton exchange membrane electrolyzer integrated hybrid multi-generation system |
| Battery-assisted low-cost hydrogen production from solar energy: Rational target setting for future technology systems |
| Bipolar plate development with additive manufacturing and protective coating for durable and high-efficiency hydrogen production |
| Comprehensive analysis and multi-objective optimization of a power and hydrogen production system based on a combination of flash-binary geothermal and PEM electrolyzer |
| Computational Modelling of the Flow Field of An Electrolyzer System using CFD |
| Construction and operation of hydrogen energy utilization system for a zero emission building |
| Control and energy efficiency of PEM water electrolyzers in renewable energy systems |
| Design analysis and tri-objective optimization of a novel integrated energy system based on two methods for hydrogen production: By using power or waste heat |
| Design and modeling of a multigeneration system driven by waste heat of a marine diesel engine |
| Design and simulation of the PV/PEM fuel cell based hybrid energy system using MATLAB/Simulink for greenhouse application |
| Design and thermodynamic assessment of a biomass gasification plant integrated with Brayton cycle and solid oxide steam electrolyzer for compressed hydrogen production |
| Design of a solar hydrogen refuelling station following the development of the first Croatian fuel cell powered bicycle to boost hydrogen urban mobility |
| Development and analysis of a novel biomass based integrated system for multigeneration with hydrogen production |
| Development and exergoeconomic evaluation of a SOFC-GT driven multi-generation system to supply residential demands: Electricity, fresh water and hydrogen |
| Development and testing of a highly efficient proton exchange membrane (PEM) electrolyzer stack |
| Development of a hybrid solar thermal system with TEG and PEM electrolyzer for hydrogen and power production |
| Development of a novel cost effective methanol electrolyzer stack with Pt-catalyzed membrane |
| Development of high pressure membraneless alkaline electrolyzer |
| Development of hybrid photovoltaic-fuel cell system for stand-alone application |
| Dimensionless approach of a polymer electrolyte membrane water electrolysis: Advanced analytical modelling |
| Dynamic behaviour and control strategy of hightemperature proton exchange membrane electrolyzer cells (HT-PEMECs) for hydrogen production |
| Dynamic hydrogen production from PV & wind direct electricity supply e Modeling and techno-economic assessment |
| Effect of catalyst distribution and structural properties of anode porous transport electrodes on the performance of anion exchange membrane water electrolysis |
| Efficient hydrogen production from aqueous methanol in a PEM electrolyzer with porous metal flow field: Influence of PTFE treatment of the anode gas diffusion layer Electrochemical hydrogen compressor: Recent progress and challenges |
| Electrochemical hydrogen production from thermochemical cycles using a proton exchange membrane electrolyzer |

Figure 2: 36 Top Papers of Interest Used in this Study

#### b. Workflow of the H2GR System

The development workflow of the H2GR system as seen in Figure 3 is broadly categorized into three distinct but inter-dependent compartments: the Domain and Decision Intelligence (DDI) compartment, the Artificial Intelligence (AI) compartment and the UI / UX compartment.

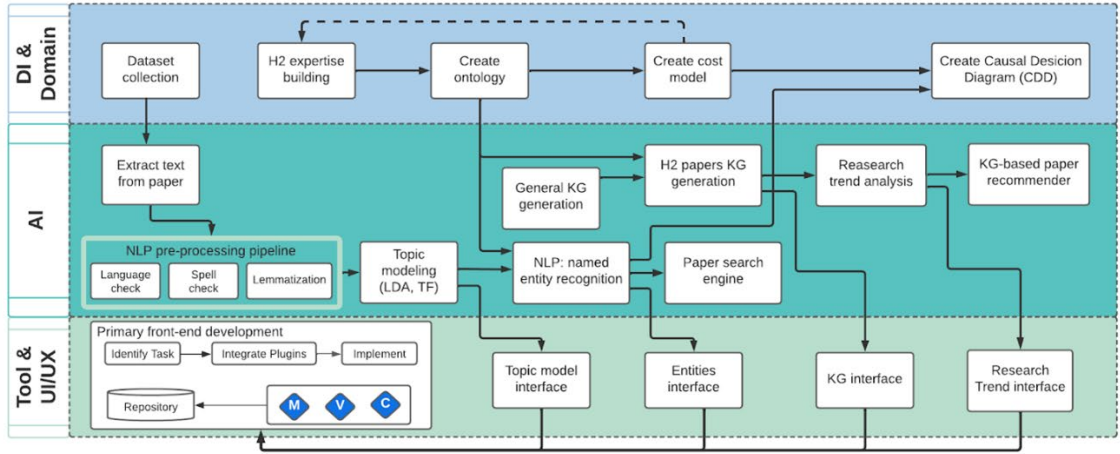

Figure 3: Workflow of the H2GR System

The major activities in the DI compartment include dataset collection, ontology and cost model creation and construction of Causal Decision Diagram (CDD). The CDD was developed to measure and visualize the variability of different components in H2 production and their expected corresponding outputs. This will assist experts and stakeholders in knowing what outcome is expected of the H2 production system. The Artificial Intelligence (AI) domain is the computational compartment which leverages on the power of state-of-the-art machine learning tools in the paper indexing, extraction, processing, and knowledge graph generation. Natural Language Processing (NLP) is the backbone of other machine learning tasks especially in the aspects of paper recommendation and research trend analysis. The Tool (UI/UX) domain is the presentation layer of this work. It presents the experts with a flexible, dynamic, and interactive user interface for tailored paper search, paper listing by recency and relevance, ontology development and optimization, topic modeling, named entities extraction, and research trend interface. It also takes care of all relevant visualizations in the system. The important components (the NLP, Ontology, KG and DI/CDD components) are discussed in the following section.

### c. System Architecture of the H2GR System

The proposed H2GR was developed using a multi-tenant architecture (MTA)[20]. In MTA, the single instance of a software serves multiple customers; each of which is a tenant given a certain level of control over the customization of the system's rules. It also enables the instances of the application to operate in a shared environment. The high-level architecture of the system is presented in Figure 4. Users/clients access the tool using any internet-enabled device. The clients' requests are served using a standard three-tier architecture, that is, the HTTP request-response protocol. The multi-tenant application is divided into three (3) distinct layers: the user interface (UI) layer, the processing layer and the data access layer, each of which interacts using a standard model-view-controller (MVC) pattern[24-26]. The user interface provides users with listings, displays, forms, diagrams, graphs and layouts by which the user interacts with the tool using event-driven approach[27, 28]. Tasks, monitored by the task manager component, are triggered depending on the user interactions and needs.

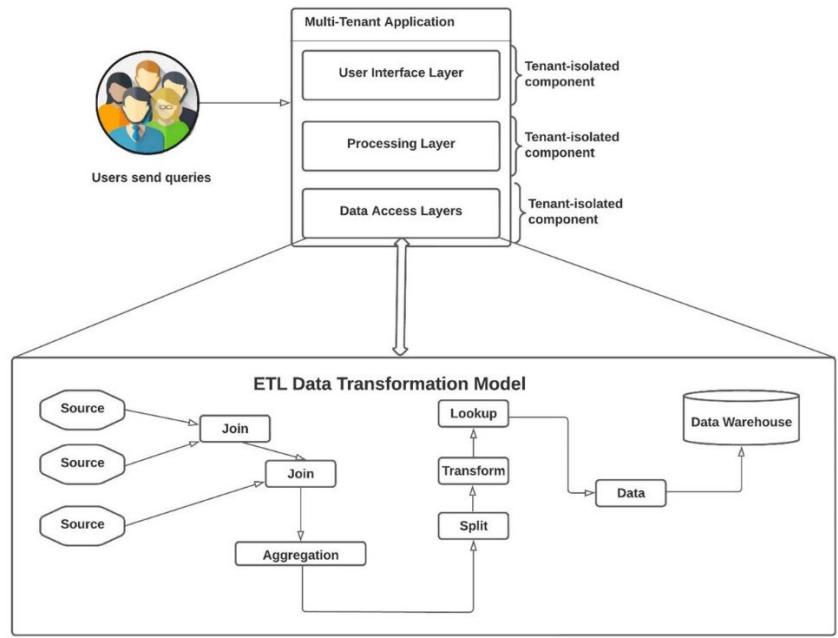

Figure 4: Multi-Tenant System Architecture of the H2GR System

Requests made to the data layers are served using the Extract-Transform-Load (ETL) data transformation model [29-31]. This model is necessary given the nature of the data processed by the H2GR. The ETL is a three-phase model in which data is extracted, transformed, and loaded into an output container. It enables our dataset to be collated and processed from one or more sources and to be outputted to one or more destinations. The "extract" component involves extraction of data from homogeneous and heterogeneous sources and in various formats including relational databases, XML, JSON, flat files, and markups from web-spidering. The "transform" component is concerned with the application of series of mapping rules or functions applied to the extracted data to prepare it for loading into the view component. The "load" component is the end point of the data access layer where data is presented in a format and structure of choice.

### d. Component-based Software Detailing of the H2GR System

The H2GR system integrates various components for achieving the tasks identified in this study in an effective AI-assisted manner. The key components in the system work together to achieve the objectives of this study and are discussed below:

i. The NLP Components: Natural Language Processing enables collection and text extraction papers, which then needs preprocessing (e.g., removal of stop words, tokenization, spelling correction, stemming, and lemmatization). Some Natural Language Understanding (NLU) tasks such as topic modeling, named entity extraction, and ontology refinement are helpful for grouping and structuring the language data. Text extraction is the process which generates the corpus, or text dataset. The methods most relevant for this work are web scraping and optical character recognition. Web scraping techniques are used to gather text from online, and typically need to be cleaned in order to remove HTML artifacts (such as resolving hyperlinks or DOM objects)[32]. Optical Character Recognition (OCR) is the process of using the image of text, such as in a PDF document, to extract a text dataset[33]. These techniques would allow the collection of quickly evolving online text, and collection of published works.

After collecting the corpus, preprocessing is the step in which the text is converted into a more usable state. A common practice is the removal of stop words (such as "the", "an") which are not considered helpful in deriving meaning from the text. What is considered a stop word will vary from one domain to another and may also change based on what is trying to be observed about the text[34]. Another common practice is spelling corrections, which is a larger problem with online text than text in publications. Then it is common to try to reduce the representations of similar concepts using either stemming or lemmatization, both of which attempt to remove grammatical modifications on words, such as tense changes, but lemmatization takes the reduction farther, reducing less common synonyms to their more common base (like "better" becoming "good")[35]. At the end of preprocessing, the text is in a cleaned and simplified state.

Named entity recognition is another technique to understand natural language, but in NERs case, you are attempting to extract references to some concept (the "entity") and there are a variety of methods to aid in this extraction[36-38]. It is a typical process in services such as search engines which need to find occurrences and relevance of words/phrases across documents. Some NLP techniques involved in this study include word tokenization, word stemming and lemmatization, topical modeling, named-entity recognition, summarization, word cloud and keyword extraction. These tasks use both linguistics and mathematics to connect the language of humans with the language of computers.

➔ *Word Tokenization:* The raw tweets after preprocessing and cleaning is broken down into smallest recognizable words and punctuations known tokens[39], the goal of which is generate the list of words which eventually is used for word cloud, summarization and sentiment analysis. The accuracy of this task is often influenced by the training vocabulary, unknown words and our-of-vocabulary (OOV) words.

➔ *Word Stemming and Lemmatization:* This transforms our tokens to its base – dictionary – form by filtering the affixation or by changing a vowel from the word[40]. Stemming and lemmatization aim to reduce the inflectional forms of a word and occasionally related derivational forms to a root form.

➔ *Topic Modeling:* Topic modeling is a technique for unsupervised categorization of the $H_2$ documents which helps to identify natural groups of words even when we are not certain what the outcome will be. This gives us a general understanding of the relationships among entities in our corpus. In this study, we used the Latent Dirichlet Allocation (LDA), a particularly popular algorithm for achieving this task[41, 42]. This is a mathematical method for finding the mixture of words associated with each topic while also determining the mixture of topics that describes each document. LDA is a generative probabilistic model of a corpus which follows a generative process for our document, $\tau$ as given in Table 1.

*Table 1: Algorithm for LDA Modeling of our Dataset*

| |
|---|
| **Input:** $H_2$ Corpus |
| **Output:** n-Gram Combinations |
| 1.     Choose $N \sim \boldsymbol{Poisson}(\boldsymbol{\xi})$. |
| 2.     Choose $\boldsymbol{\theta} \sim \boldsymbol{Dir}(\boldsymbol{\alpha})$. |
| 3.     For each of the $N$ words $w_n$: <br>      (a)   Choose a topic $z_n \sim \boldsymbol{Multinomial}(\boldsymbol{\theta})$. <br>      (b)   Choose a word $w_n$ from $p(w_n \mid z_n, \beta)$, a multinomial conditioned on the topic $z_n$. |

A $n-$dimensional Dirichlet random variable $\theta$ takes values in the ($k-1$)-simplex, that is, a $k$-vector $\theta$ lies in the $k-1$ simplex $iff$ $\theta_i \geq 0, \Sigma_{\{i=1\}}^{k}\theta_i = 1$ and has the probability density function on this simplex as given in Equation (1)

$$p(\theta|\alpha) = \frac{\Gamma(\Sigma_{\{i=1\}}^{k}\alpha_i)}{\Pi_{i=1}^{k}\Gamma(\alpha_i)} \theta_1^{\{\alpha_1-1\}} \ldots \theta_k^{\{\alpha_k-1\}} \ldots\ldots\ldots\ldots\ldots\ldots\ldots\ldots\ldots\ldots\ldots\ldots\ldots\ldots\ldots\ldots\ldots(4)$$

where $\alpha$ is a $k-$vector with component $\alpha_i > 0$, $\Gamma(x)$ is the Gamma function.

➔ *Word Cloud:* We used WordCloud, also called TagCloud to visually represent our Hydrogen papers. Tags are tokens, the importance of which is represented with font size or color as a depiction of word significance and word co-occurences. The size of each word in our WordCloud is given in Equation (2).

$$s_i = \left\lceil \frac{f_{max} \cdot (t_i - t_{min})}{t_{max} - t_{min}} \right\rceil \forall t_i > t_{min} \; else \; s_i = 1 \ldots\ldots\ldots\ldots\ldots\ldots\ldots\ldots\ldots\ldots\ldots\ldots\ldots(5)$$

where $s_i$ is display font size

$f_{max}$ is the maximum font size

$t_i$ is the count

$t_{min}$ is the minimum count

$t_{max}$ is the maximum count

Words in our WordCloud appear bigger the more often they are mentioned and are great for visualizing unstructured text data and getting insights on trends and patterns[43, 44].

ii. The Knowledge Graph Component: The term "Knowledge Graph" (KG) has been coined by Google in 2012 [45]. It is a computational tool for data visualization and interpretation. It is a way to represent the available world's knowledge and operate on them for various tasks such as information extraction, explanation, inference [46]. A KG is composed of nodes, edges, and labels to the edges. Each node represents an object, and the presence of edges signifies an existing relationship between these objects. KGs are applied in many fields such as health and life sciences, biodiversity, data integration in enterprises, or even on the web with WikiData. In the field of AI, KG are known as semantic networks and are utilized to enhance machine learning tools. KG, often, is tightly connected to the notion of ontology; combining domain information and problem-specific instances to generate KGs. Alternatively, KGs can be utilized to create domain-based embedding to AI tools, which are better input representations to train high-quality models [47].

Recently, a Hydrogen knowledge graph coined Hydro-Graph [45] has been proposed, which is a KG applied to the hydrogen literature. It is composed of a user's request, functional, and database module layers. This graph was developed to analyze the relation between papers and hot spots in the field of hydrogen. Optimally, our KG should provide insights into the state of hydrogen science, uncover relationships between research alleys, and discover promising research trends automatically rather than by hand. In turn, finding the appropriate papers with less time and effort leveraging the hydrogen taxonomy rather than high-level meta-data (e.g., citation, collaboration) and hand labeling of concepts are the strong relative characteristics of H2GR.

iii. Causal Decision Diagram / Cost Modeling Component: Following a Decision Intelligence (DI) methodology [23], our solution will be situated within the context of our users' available actions and measurable outcomes. Our software design and iterative feature selection will be done with this decision context in mind. In addition, we will construct a map of this decision context and, include user-interface elements in our deployment that reflect the decision mental-models used in hydrogen funding choices of experts and stakeholders.

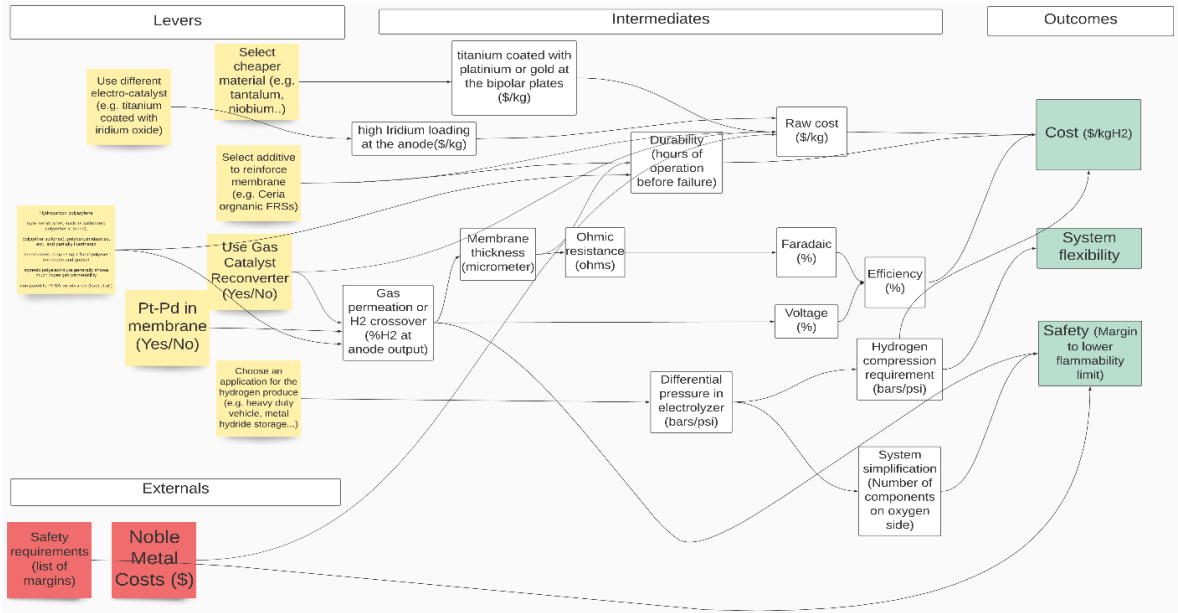

Figure 5: CDD representing the experts' decision-making opportunities to reach the "1 1 1 " goal.

The goal of a CDD as depicted in Figure 5 is to reach the "1 1 1" goal, which is measured in terms of \$/kg $H_2$ in 10 years. To do that, the stakeholders are presented with levers and sliders to allocate funding and invest in some promising research such as membrane, gold coating, Platinum loading, and Titanium contents of some components. The relations (or dependencies) between the investment to the cost reduction are visible and implementable via an observable-based software implementation. The investment to cost reduction impact, as well as a timeline on which this reduction propagates, were simulated for a real-time understanding. The relationships between the cost of component reduction and the impact on the system cost of PEM come from econometrics relationships implemented by experts in the hydrogen market. Also, the externals are the cost of raw materials of the existing and new materials utilized in the making of the PEM components. Inclusion of other elements of the decision model that affect the cost of hydrogen production and the "1 1 1" goal, typically safety-related concerns (e.g., gas permeation) and resource scarcity could be included but they were not retained for this proof-of-concept release.

iv. Ontology Component: Ontology in this work was systematically developed to improve the accuracy of our Natural Language component and build a KG to infer the most promising paper within electrolysis research. Two basic steps were followed to construct our very basic $H_2$-centric ontology: first, define the purpose and scope of the ontology. In other words, why is this ontology is being developed, and who are the end-users? This is frequently done by answering so-called competency questions [18]. Here, we developed our classical $H_2$ ontology to augment the NER in extracting information and inferring relationships from papers in the PEM water electrolysis domain. The second step is to collect knowledge about the domain and applications at hand [48], which should be reported for anyone to be able to verify the origin of the semantic relationships and adjust them if needed. Here, multiple technical reports, textbooks, and papers have been scrutinized. Overall, our main sources are [1, 13, 14, 16]. Then, the hierarchy between classes, and the relationship between concepts can be drawn within an ontological graph. For more advanced ontologies, this is when slots and corresponding facets could be defined.

There are classically three approaches to constructing ontologies [18]: top-to-bottom, bottom-up, and combination of both development processes. We opted for a top-to-bottom, where we started from the very general hydrogen production methods and narrowed it down to the catalyst in the Catalyst Coated Membrane (CCM) of the stack in a PEM electrolyzer. Regardless of the approach, it is very useful to enumerate the typical terms and verbs, in essence, the vocabulary often utilized in literature, especially if the ontological graph is fed to a NER model, thereby making it possible to identify the gaps in the knowledge domain and ensure that the concepts used are pointing to the original goal. Choosing the appropriate format for the application at hand is crucial. First, it should be implemented in a formal language to be computable by a machine and be adapted to the need of the user. Therefore, we opted for the CSV format (The ontology build will be available for the user), where the user would list the subject-predicate-object. The file is then converted into RDF/ttl format via a python script available to the user, then converted again into RDF/XML, which is readable by the UI. There are however no automatic ways or streamlined channels to evaluate the correctness and quality of our ontology. We expect the user to assess the quality of the recommendation of papers based on their search words to pinpoint if there are wrongly associated concepts. Nevertheless, at the ontology level, the connection was proofread by experts in the hydrogen field.

v. $H_2$ Paper Scoring: To build a score to recommend papers, we first contacted experts and researchers in the field of $H_2$ electrolysis to know how the quality of a paper is measured, especially as it pertains to the reputation of the author, the institution, and classically used citation-based metrics. To systematically capture the scientific relevancy of a paper, we leveraged on a graph-based score using the domain ontologies we constructed. The relevancy was a metric that most of the expert shared and we decided for this score at this early stage of the work to only consider that.

Following this, we added a PageRank algorithm applied to the domain-augmented KG in the score to reduce the bias towards reputation-based metrics, where highly established institutions and research groups would receive more credits for their work. This algorithm, invented by Larry Page [49], is a classical algorithm utilized to rank websites in the google engine. It calculates a value of a node based on the number of incoming edges at this node. Moreover, it assumes that each node spreads its importance to neighboring nodes with the same magnitude. This is mathematically represented in equation 6.

$$PR(P_i) = \gamma + (\gamma - 1)\Sigma_{P_j \in N(\mathcal{P})} P_j / C(P_j) \qquad (6)$$

where PR(X) is the score attributed to paper X, $\gamma$ is an arbitrary damping factor between 0 and 1 (we set it at 0.85), C(Pj) is the number of edges coming out of neighboring node Pj, and $\mathcal{N}(\mathcal{P})$ is the ensemble of nodes connected to $P_i$. In our analysis, we assume that all non-paper nodes point towards the title nodes, such that their page rank are all equal and constant. It might be useful, however, to have non-directed edges between papers and entity extracted, such that entities also have a non-constant score, because their score should mirror the importance of the entity itself, we are interested in within the research. We assumed no weight on the edges as well (assigned to a constant equal to 1 for visualization). This score highlights the relevancy of the papers within a particular scientific domain, rather than based on its connections to other papers in the field as it would be done via google scholar alert, which may help find untapped research direction from lower covered areas of the KG and ultimately the hydrogen research space.

## VI. Results and Discussion

### i. The Interactive User Interface

The UI/UX design of the H2GR system was built with highly interactive, highly flexible, and open-source components. It allows users to search through the system with a search term of interest (with Boolean operators such as OR, AND, EXCLUDE) and the result are typically a list of papers ranked by date and by relevancy, a list of entities identified by SpacyNER, a list of entities by the H2-Ontology, Word Cloud and a Knowledge Graph showing various entities and their relationships. Figure6 shows the landing page of the H2GR system with interactive menus for navigation. Figure 7 shows the result of a search term as displayed by the system. The graph shows the star representations of various entities related to the search term as well as their relationships. The edges are papers that connect two or more nodes, and the weight of the edges represents the amount of paper in the database that establish any two identified entities.

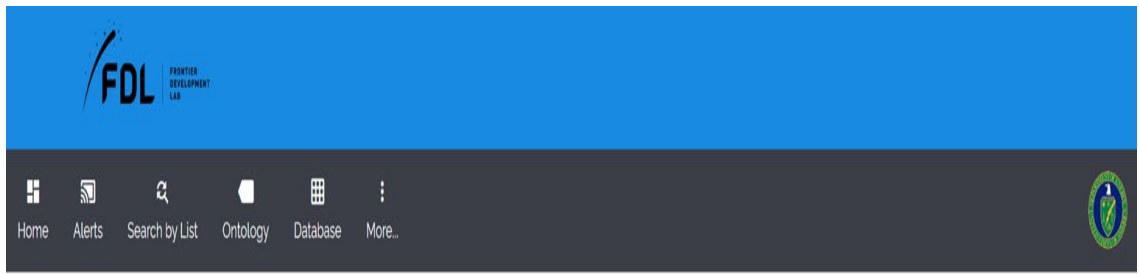

Figure 6: Landing Page of the H2GR System

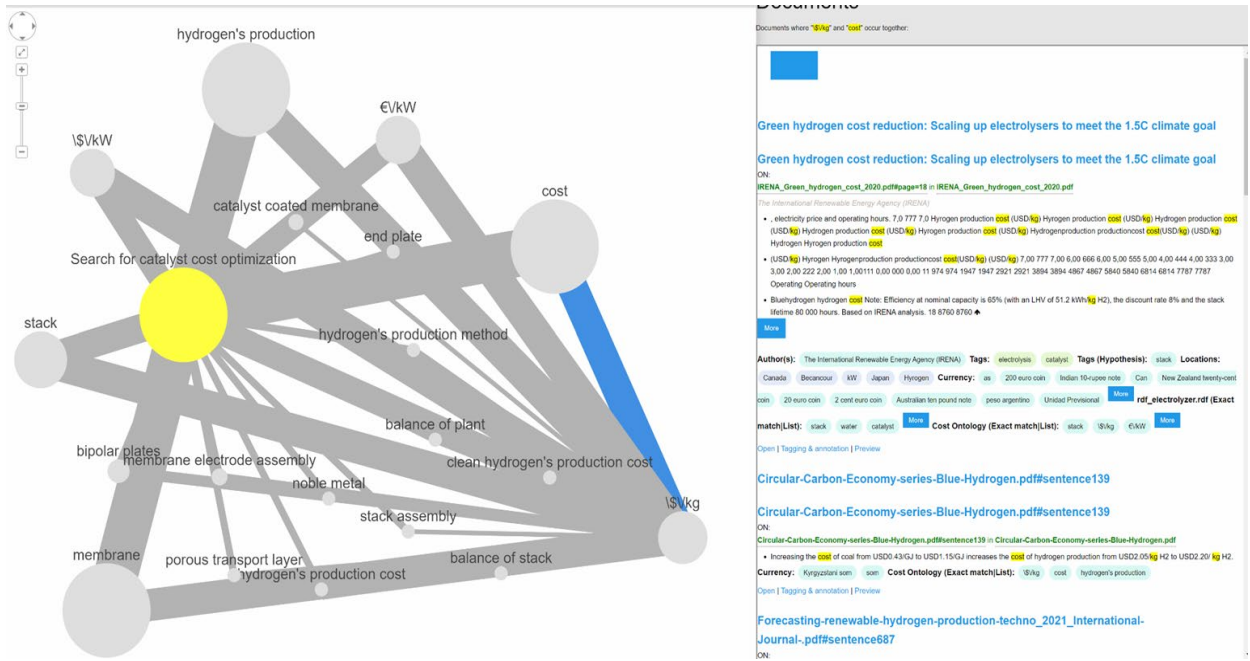

Figure 7: Search Result as Displayed in the H2GR System

### ii.     Ontology

With the preprocessed text, we applied two named entity recognition techniques: The Stanford-NER and the spaCy-NER. The Stanford-NER has some prebuild basic ontologies which are quick to implement and fairly accurate, but the only ones that were utilized in this work were the "Organizations" and the "Currency". "Organizations" ontology enabled the identification of items such as the "Department of Energy" or "Journal of Hydrogen," where as the "Currency" ontology identifies entities such as currency names (e.g., "Yen") and specific amounts "$500." However, this method is limited, so we also utilized the spaCy package's NER which utilizes the BLOOM language model and our ontologies to identify entities related to Hydrogen (extrapolated from our ontology). Once trained, it can identify entities such as "Catalyst coated membranes" or "Balance-of-plant".

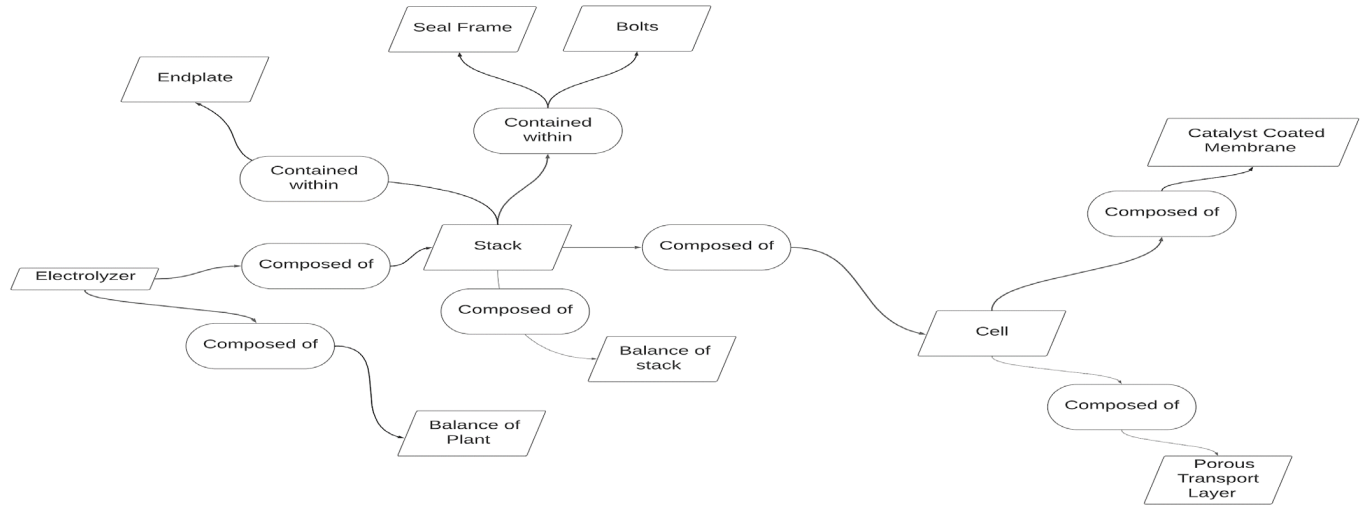

Figure 8: Sample Ontology Built with the Help of Domain Experts

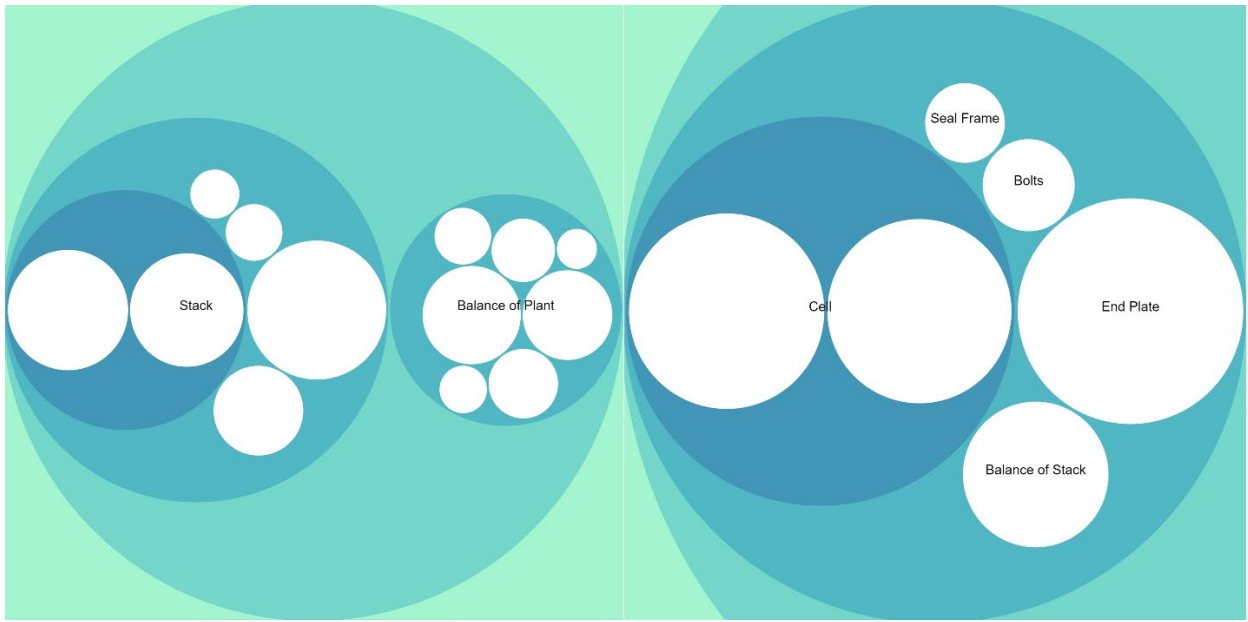

Figure 9: Bubble Plot of the H2-Ontology

Figures 8 and 9 show a sample ontology display for Hydrogen production through water electrolysis. It consists of various entities including endplate, seal frame, electrolyzer, balance of plant, balance of stack etc... and predicates,– that is, relationships – which include contained-within and composed-of. This gives more semantic meaning to the NLP and AI engines for a better machine understanding of the field.

The H2GR engine was used for several search terms to ascertain the performance of the developed system and the impact of the domain-specific ontology (H2-Ontology) was assessed. Table 2 shows the qualitative assessment of the performance of H2GR system with SpacyNER and the H2-Ontology developed in this study when a sample search key was fed into the system. The number of papers returned, number of entities returned and the category of meaningful entities when the H2-Ontology was applied surpasses the SpacyNER by a great margin.

Table 2: Performance of H2GR with Spacy H2-Ontology when searching for 'Balance of Plant'

| Metric | SpacyNER | H2-Ontology |
| --- | --- | --- |
| Number of Papers Returned | 567 | 1163 |
| Number of Meaningful Entities Returned | 1008 | 12846 |
| Top Meaningful Entities Returned | Authors, Organizations, Plant, Power, Country, Stack | Balance of Plant, Stack, Ccell, Catalyst-coated Membrane, Electrolyzer, Balance of Stack, Cost |

### iii. Cost Modeling / Causal Decision Diagram

In this study, a cost modelling environment with an intuitive nested tree and tornado chart was developed to represent the impact of the cost components in the Hydrogen electrolyzer stack. This helps experts and stakeholders to ideate around the decision to make evidence-based research funding allocations.

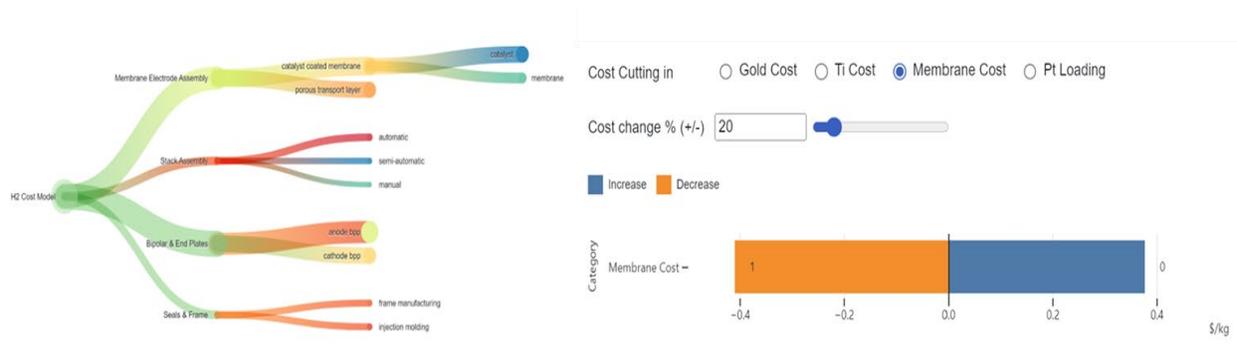

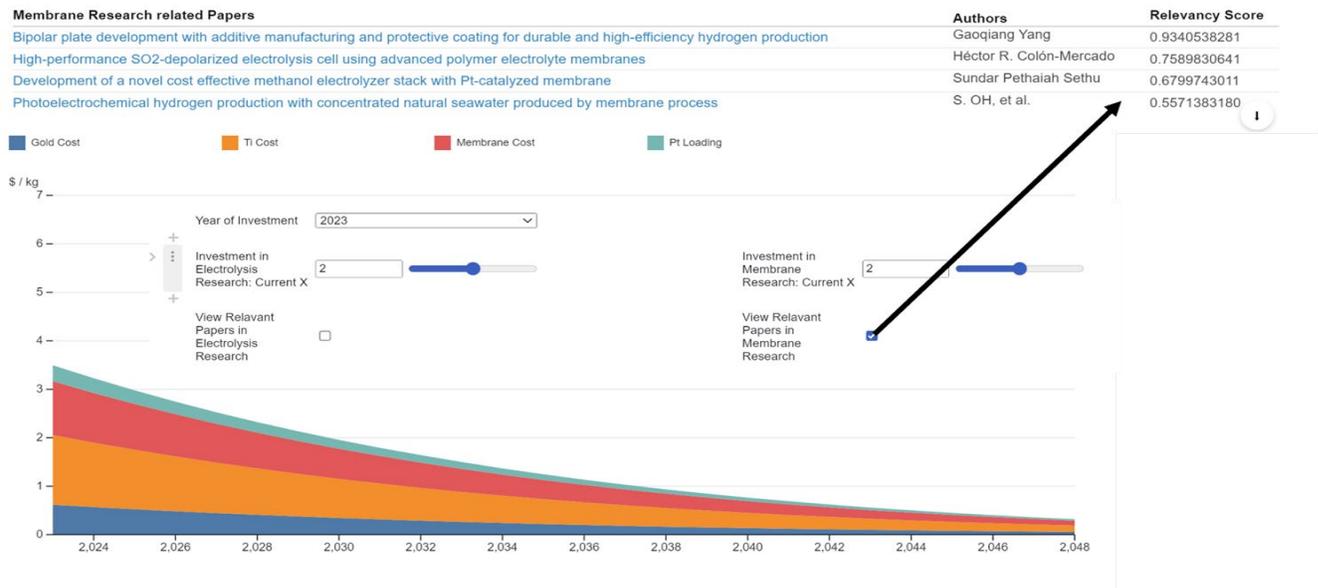

Figure 10: Causal Decision Diagram / Cost Modeling of Membrane-Related Research Papers

Figure 10 shows the decision modeling platform containing simulation environment for membrane-related research papers. This will help to establish dependencies and relationships among the components identified in membrane papers. It also displays papers which relates to the costs of these components with a relevancy score relative to the component in question. The investments are represented by the lever. Years at which we are investing in also matters. The CDD is connected to the UI via the clickable boxes. Then a list of papers relevant to the topic of the cost parameter is given and ranked via a Learning-To-Rank (LTR) - based algorithm.

iv. Topic Modeling

The important components of the NLP pipeline are the NLP preprocessing, the topic modeling, and the named entity recognition which is then combined with meta-data to create the KG. With the preprocessed text, we applied LDA to uncover topics at two levels of granularity: the page, and the paragraph. Figure 11 depicts the perplexity levels for the increasing number of topics. The lower the perplexity, the better of a fit the topic model. A common technique for selecting the number of topics to use in a topic model is called the "elbow technique" in which you pick the point at which the perplexity begins to drop less drastically (as perplexity tends to continually drop as you add more topics). This technique would select seven topics for the Document level topic model, and six topics for the Page level topic model. Descriptions of those topic models can be seen in Figure 12 and Figure 13, respectively.

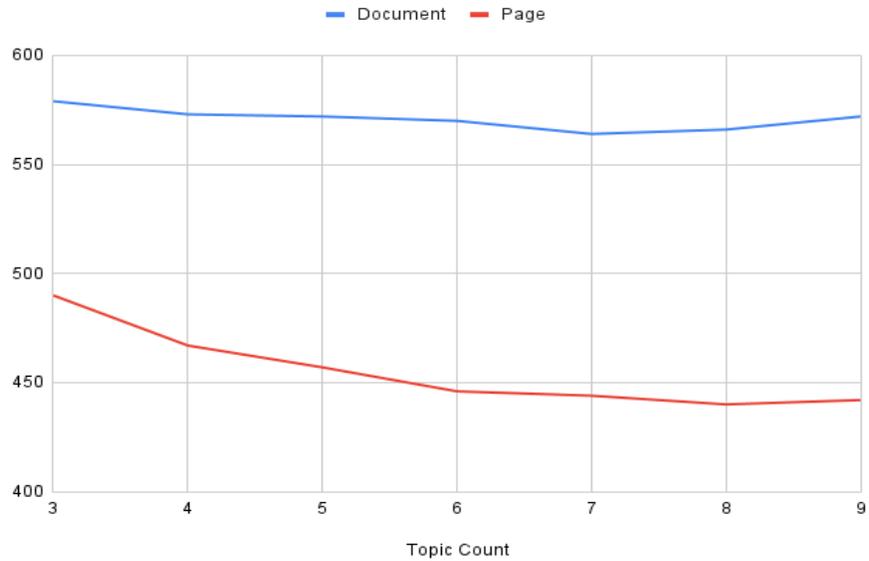

Figure 11: The perplexity of LDA topic models at varied topic amounts.

On the average, the Page level perplexity is quite lower than the Document level perplexity. This means that LDA was able to find better fitting topic models at the Page level, likely due to high redundancy across documents that is substantially reduced when you focus on pages.

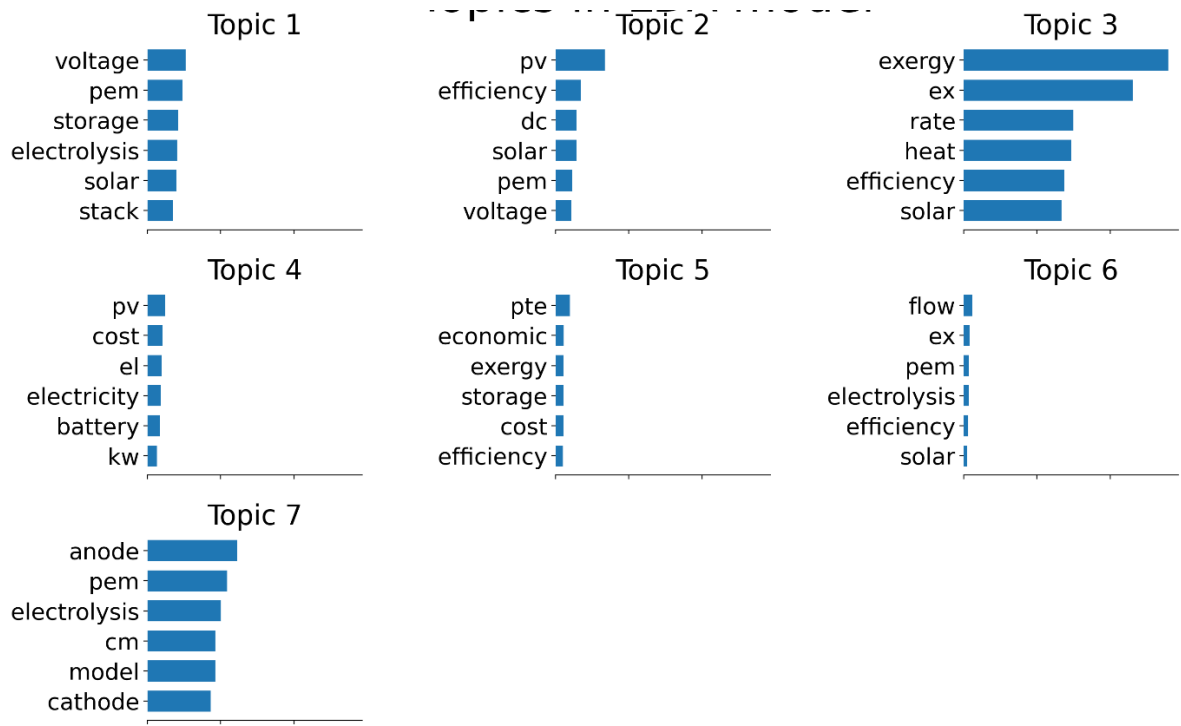

Figure 12: The set of topics with seven topics when entire publications are used as the documents

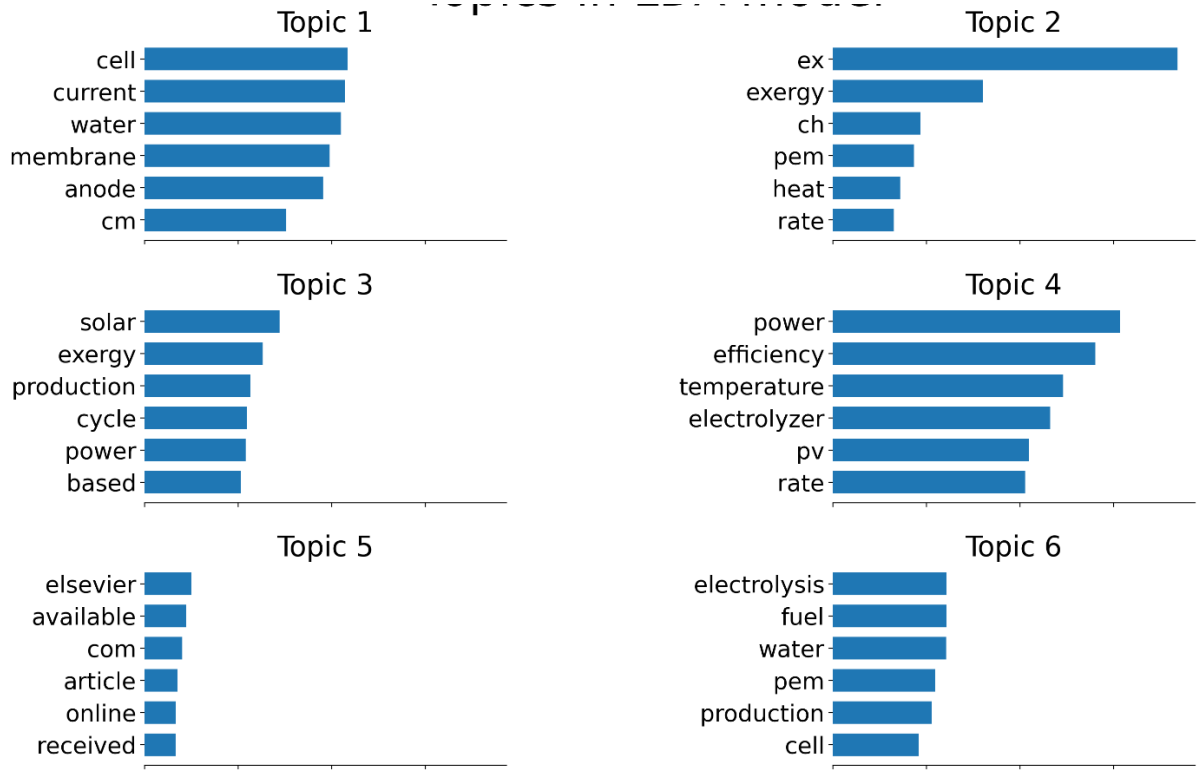

Figure 13: The set of topics with six topics when pages are used as the documents

These two topic models can be used in concert to attempt to categorize a document and characterize the topic distribution and frequency across the document's pages. The topic model was also integrated into the H2GR to help a user focus on certain pages of relevant publications.

v. Scoring Publications in Knowledge Graph

We applied standard methods to ranking publications in the H2GR system. This was done by generating a score which highlights the relevancy of the papers within a scientific domain rather than based on just its connections to other papers in the fieldFigure 14 showcases an example of scores obtained within a KG constructed with meta-data.

Figure 14: Page Rank algorithm applied to 36 papers ranked in a knowledge graph constructed with meta-data: Year of publication, author, citation, publication, url, citation over time, and timestamp. Here the paper "A solid oxide

membrane electrolyzer for production of hydrogen and syn-gas from steam and hydrocarbon waste in a single step" (highlighted in red) receives the highest score.

However, we applied the exact same dataset with two other ontologies as well as combined the meta-data and these ontologies. The two ontologies are the domain science and cost. The domain science are all the terms related to hydrogen production with more emphasize on PEM electrolyzer, while domain cost are all the concepts related to cost within the hydrogen production domain (e.g., $/kWh). The best papers provided each time are given in Table 3:

Table 3: Papers with the highest score obtained with different combination of ontologies.

| Ontologies | Paper Title |
|---|---|
| Meta-data | A solid oxide membrane electrolyzer for production of hydrogen and syn-gas from steam and hydrocarbon waste in a single step |
| Domain Science | 1D two-phase, non-isothermal modeling of a proton exchange membrane water electrolyzer: An optimization perspective |
| Domain Cost | Development and testing of a highly efficient proton exchange membrane (PEM) electrolyzer stack |
| Domain Science and Cost | Battery-assisted low-cost hydrogen production from solar energy: Rational target setting for future technology systems |
| Combination of the Three | Development and testing of a highly efficient proton exchange membrane (PEM) electrolyzer stack |

It is interesting to note in parallel with Table 4, and as pictorially represented in Figure 15, that the score for the paper "1D two-phase, non-isothermal modeling of a proton exchange membrane water electrolyzer: An optimization perspective" when only looking at meta-data, is rather low. This can be explained because this paper is from February 2022 and has no citation. However, within the domain science and cost, its score is among the top. On the other hand, the paper "A solid oxide membrane electrolyzer for production of hydrogen and syn-gas from steam and hydrocarbon waste in a single step", which has the highest score within the meta-data but a low score among the other domain ontologies, has among the most amount of citation and has been published much earlier in 2011. As a result, this highlights that if a domain ontology was not utilized to build the KG, the recommender could miss some important papers relevant within the research of interest. Reputation-based metrics can sometime therefore be insufficient.

Table 4: Score of the papers recorded in Table 3. Paper 1 is "A solid oxide membrane electrolyzer for production of hydrogen and syn-gas from steam and hydrocarbon waste in a single step", Paper 2 is "1D two-phase, non-isothermal modeling of a proton exchange membrane water electrolyzer: An optimization perspective", Paper 3 is "Development and testing of a highly efficient proton exchange membrane (PEM) electrolyzer stack", and Paper 4 is "Battery-assisted low-cost hydrogen production from solar energy: Rational target setting for future technology systems". The magnitudes depend on equation (6).

| Paper # | Meta-Data | Domain Science | Domain Cost | Domain Science and Cost | Combination of the Three |
|---|---|---|---|---|---|
| 1 | 0.0157 | 0.0121 | 0.0121 | 0.009 | 0.0124 |
| 2 | 0.0106 | 0.0260 | 0.0227 | 0.0213 | 0.0137 |
| 3 | 0.0157 | 0.0224 | 0.0232 | 0.0225 | 0.0179 |
| 4 | 0.0142 | 0.0174 | 0.0306 | 0.0236 | 0.0173 |

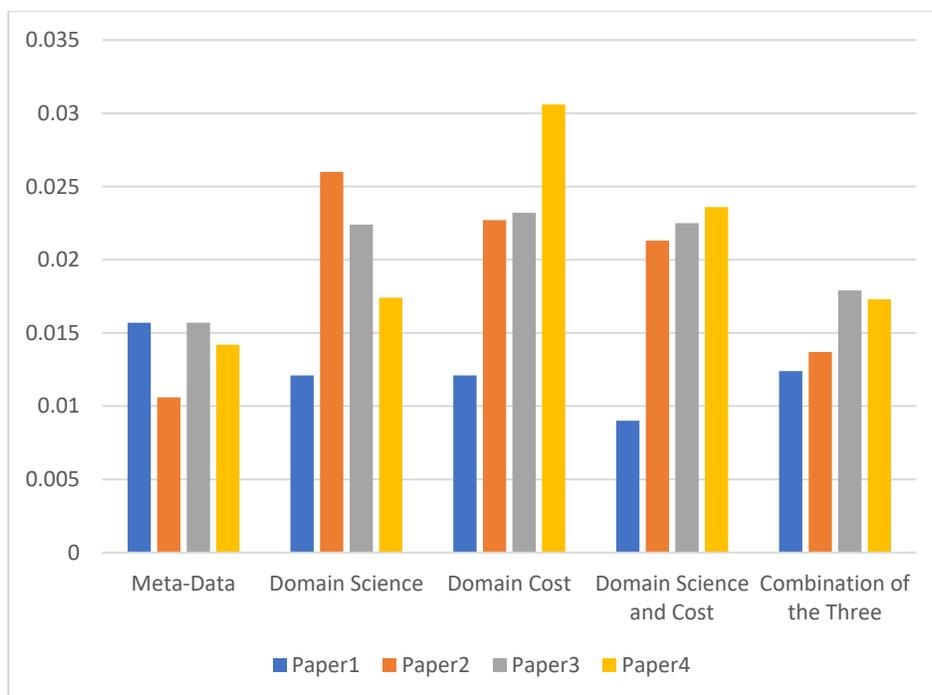

Figure 15: Relevancy-Based Ranking of Papers in Table 3

This score is integrated with reputation-based metrics alluded above. This can be adopted as it mimics how the research community is interacting with an existing and new corpus of literature, thereby, making it suitable to detect promising research directions.

## VII.     Conclusion and Future Work

Hydrogen energy is an extremely promising factor to help decarbonize the economy, but methods for producing it with zero-carbon in a cost-effective way are still in an area of open research. To achieve this, funding decisions must be made efficiently in a timely manner, as the DOE pledge to reduce its cost to 1$/kg by the end of the decade. Therefore, a pipeline that scrapes and indexes relevant hydrogen papers from relevant journals was developed. We applied NLP for text and entity extraction, the establishment of relationships between papers, and Named Entity Recognition (NER) as well as paper ranking using relevance score to suggest promising papers. We also used a human-in-the-loop strategy by working with some Hydrogen experts in developing ontologies for a domain semantic-augmented NLP process and the impact of the ontology on the system was measured. The user is also able to interact with a Causal Decision Diagram (CDD) to make optimal research funding allocation decisions. Authors have made significant progress in representing and organizing the hydrogen research space. We demonstrated that using an ontology to build a KG changes the relevance of papers and highlight papers that would not be seen by using traditional citation-based metrics. Initial attempts at linking that space to economic models (through Decision Intelligence) have also been attempted.

In this study, we proposed H2GR, a Hydrogen-research Golden Retrieval System for efficient Hydrogen research discovery and analysis. It presents experts and stakeholders with a robust Hydrogen papers database and tools for a quicker trend and quality analyses towards an evidence-based Hydrogen research grantsmanship. This will help experts to sift through papers and identify new trends, patterns, where to invest how much, among others. Arguably, this method can seamlessly be applicable to any field as long as the ontology is adapted to the targeted domain. Experts who tested the work liked the possibility to refine search words by adding tags and notes presented by the H2GR system. They also liked the graphical representation which allows for visual analysis of what concepts are contained within papers.

In spite of the many advantages of the developed system, there are still many avenues for improvement regarding the ontology, the data incorporated in the search engine, and the training of the different algorithms. They include:

> Streamlining the implementation and interaction of the ontology via a user-friendly CSV interface (e.g., interactive visualizer, color coding depending on properties and hierarchy, automatic rdf/xml conversion...) and extend it with the help of experts. Expand the RDF implementation to give more information for the ontology (e.g., axioms and facets) and integrate existing general ontologies. Uncover and infer new axioms, concepts, and relations with the help of automatic or semi-automatic tools.

> Making the cost modeling dynamic and input true data to the CDD with the help of an expert. Inclusion of other elements of the decision model that affect the cost of hydrogen production and the "1 1 1" goal, typically safety-related concerns (e.g., gas permeation), resource scarcity concerns (e.g.: PGM), and tax credits.

> Use more advanced scoring metrics including the customer ways of looking for relevant research, and the impact of new research on the cost (which is the penultimate goal of such work). In this prospect, extract information from papers (e.g., what is the cost of material X) rather than entities.

> Implementing Graph learning (e.g., K-mean clustering) for improved interaction and visualization.

> Use time series prediction to predict the citation of the papers. Training a GNN to turn the problem into a classification problem using the KG generated as input embedding will be given attention in the nearest future.

**Acknowledgement**

This work has been enabled by the Frontier Development Lab (FDL.ai). FDL USA is a collaboration between several government agencies, Department of Energy (DOE), National Aeronautics and Space Administration (NASA), and U.S. Geological Survey (USGS), SETI Institute, and Trillium Technologies Inc., in partnership with private industry and academia. This public/private partnership ensures that the latest tools and techniques in Artificial Intelligence (AI) and Machine Learning (ML) are applied to basic research priorities in support of science and exploration of material concerns to humankind.

We would like to wholeheartedly thanks everyone who contributed to the success of this work and who are not referred to as co-authors. Anne Marie Esposito and Marika Wieliczko at the Department of Energy for their conception and formulation of the challenge approach, as well as providing and collecting expert insight and data sources, and provide guidance on model limitations and improvement. John Petrovic for his expert insight into the hydrogen water electrolysis domain. And lastly Sabine Scholle and Arpitha Malavalli who spent their free time implementing the DI and cost model visualization and interaction tools.